\documentstyle[sprocl]{article}

\def\Journal#1#2#3#4{{#1} {\bf #2}, #3 (#4)}

\begin{document}

\title{THE SIGNATURE OF DARK MATTER IN QUASAR LIGHT CURVES} 

\author{M. R. S. HAWKINS}

\address{Royal Observatory, Blackford Hill, Edinburgh EH9 3HJ,
 Scotland\\ E-mail: mrsh@roe.ac.uk} 

\maketitle\abstracts{ 
The idea that dark matter in the form of small compact bodies
(most plausibly planetary mass primordial black holes) betrays
its presence by the microlensing of quasars, has been stimulated
by statistical studies of quasar light curves and gravitationally
lensed quasar systems.  In this paper we review this evidence, and
discuss the significance of the recent detection of a planetary mass
microlensing event in the Galaxy by the MACHO group.  We also look
in more detail at the light curves themselves, and make the case
that many show the characteristic morphology of caustic crossing
events associated with a large optical depth of lenses.  This
supports the idea that a large fraction of the critical density is
in the form of planetary mass compact bodies.  The nature of such
bodies is discussed, and it is argued that current work on the QCD
phase transition suggests they are primordial black holes, although
it is just possible that they are compact gas clouds.
}

\section{Introduction}

There are a number of lines of argument which suggest that a
substantial amount of dark matter may exist in the form of compact
bodies.  Although many mass ranges are trivially ruled out from
the gross effects which would be observed, the range from
$10^{-6}$ to one $M_{\odot}$ is still largely unconstrained
outside the Galaxy.  Within the Galaxy, the MACHO and EROS projects
suggest a complicated picture which has still not been finalised.
In this paper we review the evidence favouring a significant
population of compact bodies in the subsolar and planetary mass
range.  This comes largely from an analysis of quasar light
curves, both single objects and gravitationally lensed systems,
to test the hypothesis that variation is caused by the microlensing
effects of a large population of small bodies.  It is possible
that one such body has recently been detected by the MACHO group
in the Galaxy.  If quasar microlensing is a general phenomenon,
the mass of the lenses can be estimated at $\sim 10^{-3}_{\odot}$.
Further arguments suggest that the lenses are most plausibly
primordial black holes and make up a large fraction of the
cosmological critical density.

\section{Observational Evidence for Microlensing}

\subsection{Statistical Analysis of Quasar Light Curves}

Until a few years ago it was generally assumed that observed
variations in quasar brightness where caused by intrinsic changes
in the flux output of the quasar.  For short variations of a few
months, the observed timescales broadly match theoretical
expectations, although precise mechanisms for the brightness changes
are still hard to specify.  Long term variations of a few years,
which appear to dominate quasar light curves, are much harder to
explain as intrinsic variation, not only because of the timescale,
but also other statistical properties.  An alternative proposal
by Hawkins~\cite{mh1,mh2} was that the long term variation was
caused by microlensing.  A population of sub-stellar or planetary
mass bodies along the line of sight would produce light curves
similar to those observed, and the amount of material required
would be a large fraction of the cosmological critical density.
By measuring the timescale of variation and making some assumptions
about the kinematics of the microlensing bodies it is possible to
estimate their characteristic mass.  This turns out to be around
the mass of Jupiter, with an uncertainty of at least an order of
magnitude.

Variation caused by microlensing has a number of statistical
properties which can be tested by analysis of a large sample of
quasar light curves.  The tests include measures of chromatic
variation, statistical symmetry, luminosity effects and time
dilation (intrinsic variation should show an increase of timescale
with redshift).  Results of these tests and others have been
published~\cite{mh2}, and in all cases are consistent with the
predictions of microlensing.  While such an approach cannot easily
rule out the unconstrained possibilities of intrinsic variation,
the results are hard to model on that basis~\cite{ht}.

\subsection{Microlensing in Multiply Lensed Quasar Systems}

Quasars in multiply lensed systems are known to vary in two
distinct modes.  The best known of these is when all images vary
in the same way with identical light curves, but with time
differences for each image.  This is well understood
to be the effect of intrinsic variation in the parent quasar,
which due to small differences in light travel time results in
temporal displacement of the light curves.  However, in all
quasar systems where there is sufficiently comprehensive
monitoring, additional variation in individual images is also
seen.  There is general consensus that this has to be the effect
of microlensing, although the nature and location of the
microlensing bodies is still open to debate.

One might argue that since microlensing is seen in every quasar
system, it is a generic property of all quasars, and is a
manifestation of the effect seen in the statistical monitoring
of samples of single quasars.  The problem with this line of
argument is that macrolensed quasars inevitably have a massive
galaxy close to the line of sight which may be the source of the
microlensing.  Thus lensed quasars may not be representative of
the quasar population as a whole.  In fact this argument is not
as secure as it might seem.  For a high probability of microlensing
there must be approaching a critical optical depth to lensing
in microlensing bodies along the line of sight.  This implies that
the galaxy is largely made up of microlensing bodies, and that they
make up the dark matter.  The microlensing cannot be caused by
normal stars which only make up a few percent of the galaxy's
mass.  The only alternative is that the microlensing is taking
place more generally along the line of sight as for other quasars,
so either way we must conclude that the dark matter is in the form
of compact bodies.

\subsection{Detection of Planetary Mass MACHO's in the Galaxy}

If dark matter is in the form of compact bodies, one might plausibly
expect to detect them in the halo of the Galaxy.  It is well known
that the MACHO and EROS projects claim to have detected a population
of compact bodies sufficient to make up around half the mass of the
halo, and it is tempting to associate them with bodies responsible
for quasar microlensing.  The problem with this is that the MACHO
mass is estimated at around half a solar mass, whereas microlensing
of quasars suggests a mass of $\sim 10^{-3}M_{\odot}$.  At the moment
it seems that these two mass estimates cannot be made compatible.

Until recently no planetary mass bodies had been detected by the
MACHO project, which they claim implies an upper limit of 20\% of the
mass of the halo in such objects~\cite{ca}.  However, in the Galactic
bulge a short timescale event was recently observed, with an
estimated mass about twice that of Jupiter.  This single detection
taken at face value implies a large mass density of such objects,
and a density enhancement in the bulge consistent with their
identification as the dark matter component~\cite{mh3}.  If this
is a detection of one of the objects responsible for quasar
microlensing then the limits for the halo, if correct, imply that
the mass distribution is flatter than might be expected.

\subsection{Caustic Crossings in Quasar Light Curves}

Although statistical tests of quasar light curves are important
probes of the nature of the variation, and can tightly constrain
models of both intrinsic variability and microlensing, they do
not make use of the detailed information available in the
observations.  When quasars are observed through a large optical
depth for microlensing, the lenses combine non-linearly to produce
caustic patterns.  This effect has been simulated numerically
and analytically by several groups, and it is found that
characteristic double spiked patterns are formed in the quasar
light curves as the source traverses the caustics.

Examples of such caustic crossings have recently been looked for
in a large sample of quasar light curves~\cite{mh4}.  The
double spiked structure appears to be a feature of many of the
light curves, but their interpretation as caustic crossings seems
inconclusive.  However two light curves, illustrated in Fig. 1,
stand out as being very plausible caustic crossing events.  
\begin{figure}
\includegraphics{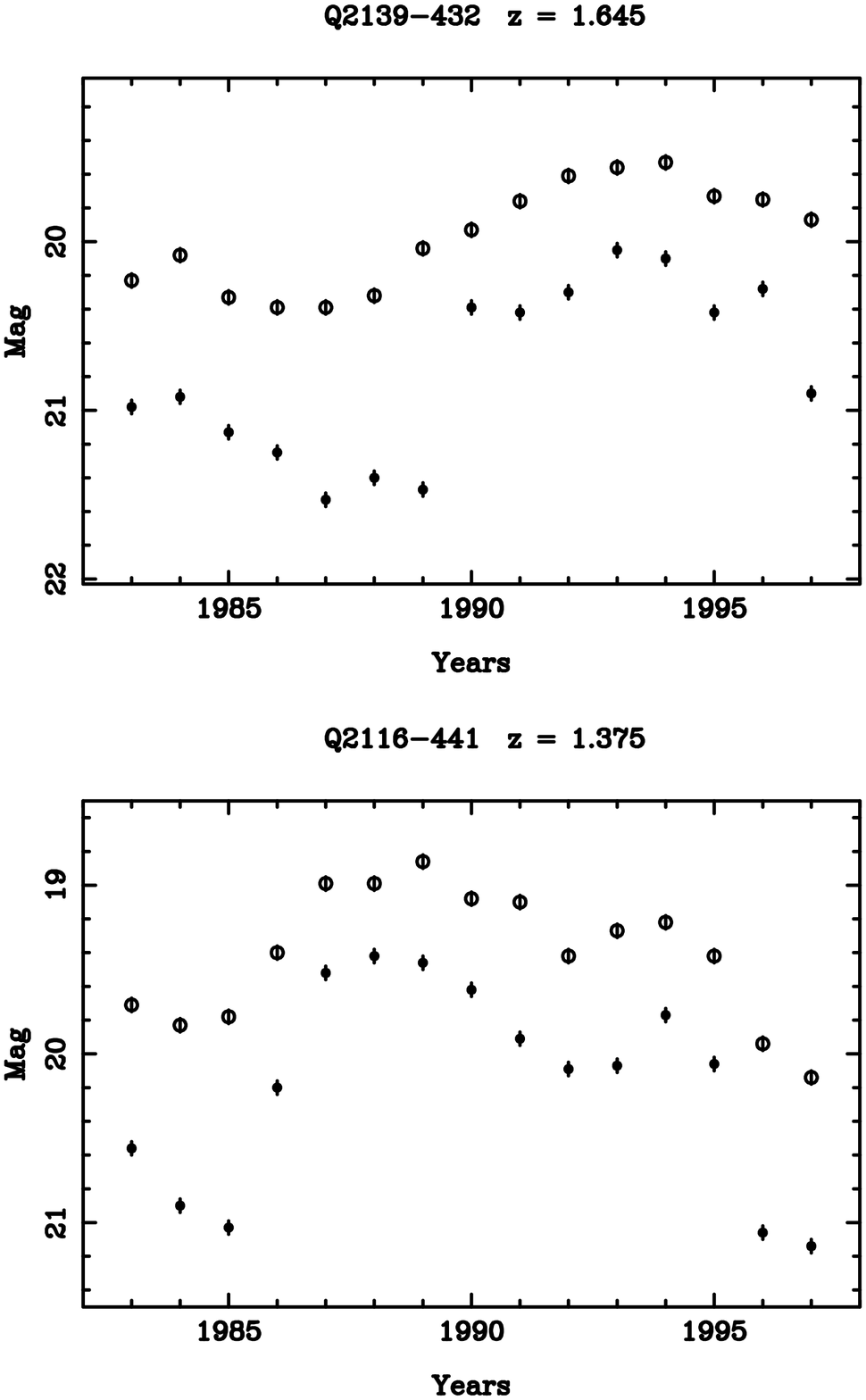}
\vspace{16cm}
\caption[]{
      Light curves for two quasars showing all the characteristics
      expected of caustic crossing events.  Filled and open circles
      are blue and red passband measures respectively. Error bars
      are based on measured photometric errors.
}
\end{figure}
In the light curve in the top panel the blue light rises very
rapidly by more than a magnitude to a cusp like feature; the red
light actually starts to increase three years earlier, but does
so smoothly, and does not achieve the same amplitude as the blue.
After six years a similar pattern in reverse appears to occur.
In the bottom panel a somewat similar picture is seen.  It is very
hard to think of a plausible model for this behaviour in terms of
intrinsic variation, although microlensing provides a straightforward
explanation.  The blue light would come predominantly from a blue
compact nucleus of an accretion disk, essentially unresolved by the
lenses.  It would thus show the characteristic features of caustic
crossing events.  The red light on the other hand would come mostly
from the much larger area of the whole accretion disk.  This would
be resolved by the lenses; the magnification would commence earlier
but never achieve the same amplitude as the blue, and there would
be no sign of caustic crossing features.

\section{Theoretical Constraints}

The arguments in the previous section all support the idea of a
population of compact bodies around the mass of Jupiter sufficient
to make up a significant fraction of the cosmological critical
density.  It is just possible that they could be baryonic.  By
pushing baryon synthesis constraints to their limit with
inhomogeneous mucleosynthesis, and reducing the required optical
depth to microlensing to the minimum that will produce the observed
level of variation one might be able to compromise on a
cosmological density $\Omega = 0.3$.  The idea that the bodies
could be ordinairy planets seems completely implausible, but
there is the interesting possibility that they could be planetary
mass gas clouds~\cite{ww}.  This solution has the advantage that
it explains why it has been so hard to detect the bodies in the halo,
as these gas clouds would be too large to micolens LMC stars.

An alternative possibility which we have argued in earlier papers
is that the microlensing bodies are primordial black holes.
Far fetched though this idea might sound, it provides a plausible
explanation of the observations, is consistent with baryon
synthesis constraints, and fits in well with current work on
black hole formation in the QCD phase transition~\cite{nj}.  The
black holes would behave as cold dark matter and make up
sufficient mass to account for dark matter on a large scale.

\section{Discussion and Conclusions}

In this paper we have reviewed the evidence implying that quasars
are being microlensed by a population of Jupiter mass bodies.
Each line of argument is probably not conclusive in itself, but
taken together they provide a structure which is hard to circumvent.
There seems to be no doubt that at least some quasars are being
microlensed; the residual question is how widespread is the
phenomenon.  This in turn directly relates to the amount of dark
matter which can be made up of the microlensing bodies.  At present
the mass density would appear to be close to the cosmological
critical density.  There is also some considerable uncertainty
in the characteristic mass of the bodies.  The best estimate at
the moment seems to be around $10^{-3}M_{\odot}$ but this could
be out by at least an order of magnitude either way.

If one accepts the strong arguments that dark matter must at least
in part be in the form of non-baryonic material, then the two most
plausible possibilities seem to be some sort of SUSY particle, or
primordial black holes.  The main problem with the idea of SUSY
particles is that at the moment their properties are largely
unconstrained, and they are close to being undetectable.  Primordial
black holes on the other hand can be detected in a predictable
way, and indeed as argued in this paper, this may already have
happened.  The circumstances in which they might form are
becoming better understood, as well as the mass function.
The only other possibility is that the dark matter is baryonic,
and in this case the best candidates would appear to be compact
planetary mass gas clouds.

\end{document}